# Fabrication and characterization of single-crystalline CoSn ($10\bar{1}0$) kagome metal thin films for interconnect applications: structure and anisotropic electrical resistivity

Tomoya Nakatani,[1,*] Nattamon Suwannaharn,[2] and Taisuke T. Sasaki[2]

[1]Research Center for Magnetic and Spintronic Materials, National Institute for Materials Science, 1-2-1, Senen, Tsukuba, Ibaraki 305-0047, Japan.

[1]Research Center for Structural Materials, National Institute for Materials Science, 1-2-1, Senen, Tsukuba, Ibaraki 305-0047, Japan.

*Corresponding author's email address: nakatai.tomoya@nims.go.jp

**ABSTRACT**

CoSn kagome metal is a pseudo-one-dimensional electronic conductor, exhibiting low resistivity ($\rho$) along the [0001] direction ($c$-axis) and significantly higher $\rho$ along other crystallographic directions. Such anisotropic conduction is expected to mitigate resistivity increases in narrow interconnect wires at advanced semiconductor technology process nodes, making CoSn a promising candidate for future interconnect applications. In this study, CoSn thin films were fabricated by magnetron sputtering, and their resistivity anisotropy was investigated with respect to crystallographic orientation. Epitaxial growth of single-crystalline CoSn ($10\bar{1}0$) films was achieved on a Ru ($10\bar{1}0$) buffer layer at deposition temperatures above 350 °C. The CoSn films exhibited relatively low $\rho$ along [0001], reaching 13 $\mu\Omega$ cm for films thicker than 50 nm, and an approximately tenfold anisotropy of $\rho$ between [0001] and [$2\bar{1}\bar{1}0$] ($a$-axis), consistent with previous reports on bulk CoSn single crystals. However, the CoSn($10\bar{1}0$) surface exhibited pronounced roughness, attributed to three-dimensional crystal growth during sputtering, which hinders accurate evaluation of the thickness dependence of resistivity. Scanning transmission electron microscopy revealed the growth of a CoSn ($10\bar{1}0$) single-crystal with ($11\bar{2}0$) and ($01\bar{1}0$) side wall facets, as well as domain boundaries within the films. These results highlight both the potential and challenges of employing CoSn kagome metal in future interconnect technologies.

---

As the physical dimensions of semiconductor devices shrink, the thickness and width of the metallic interconnect wires connecting transistors decrease. This increases the electrical resistance of the interconnects, which limits the performance of devices.[1–3] Although copper (Cu) is one of the most conductive metals and has been used for interconnects for nearly three decades, the resistivity of Cu thin films and wires significantly increases as their thickness and width decrease, which is caused by the scattering of conduction electrons at the surface/interface of the interconnects with small dimensions due to the approximately isotropic Fermi surface and the long electron mean free path ($\lambda$) (~39 nm at room temperature (RT)[4]) of Cu. Following the guideline proposed by Gall that $\rho_0\lambda$, where $\rho_0$ is the bulk resistivity, is a figure of merit for interconnect materials,[5] various metals and alloys (intermetallic compounds) have been investigated, such as Ru, Mo, Rh, Ir,[6] W,[7] NiAl,[8,9] RuAl,[10,11] and CuAl$_2$.[12,13]

In addition, materials with a strong dependence of resistivity on their crystallographic orientations are promising for future interconnect applications.[14,15] Such an anisotropic electronic conduction arises from anisotropic Fermi surfaces, which leads to smaller size dependence of resistivity compared to the materials with isotropic Fermi surfaces. Kumar *et al.* proposed several materials with anisotropic electronic conduction based on first-principles material screening.[14] Delafossites, such as $PtCoO_2$ and $PdCoO_2$, are pseudo two-dimensional conductors with low resistivity in the *c*-plane of $\rho$ = 2.1 and 2.6 µΩ cm in bulk at RT,[16] respectively, and $\rho$ down to 4.21 and 3.49 µΩ cm in thin films.[17,18] Pseudo-one-dimensional electronic conductors of CoSn, $YCo_3B_2$, and OsRu have been proposed as promising candidates for interconnect materials. Of these three candidates, CoSn may be the most suitable for the mass production process of the semiconductor devices due to its relatively low material cost.

CoSn is a kagome metal, an intermetallic compound with the B35 structure (hP6, space group P6/mmm, No. 191) with $a = b$ = 0.5279 nm, $c$ = 0.4260 nm, $\alpha = \beta$ = 90°, and $\gamma$ = 120° as depicted in Fig. 1(a).[19] CoSn shows relatively low $\rho$ of 3–7 µΩ cm along the [0001] *c*-axis, while having much higher $\rho$ of >100 µΩ cm in the (0001) *c*-plane.[20–24] This makes CoSn promising for the interconnect application. Such an anisotropic electronic conduction in CoSn derives from the "flat-band", in which the mobility of conduction electron is nearly frozen in the kagome plane (*c*-plane) due to a large effective mass. While many studies of the physical properties of CoSn using bulk single crystals have been reported,[20–24] reports of CoSn thin films are still limited. Thapaliya et al.[25] and Ikawa and Fujiwara[26] fabricated single-crystalline CoSn (0001) films on $Al_2O_3$ (0001) substrates with Pt/Ru and Co buffer layers, respectively. Cheng et al.[27] fabricated single-crystalline CoSn (0001) films on a 4H-SiC (0001) substrate. The reported values of $\rho$ in-plane (*c*-plane) were 139 and 192 µΩ cm, respectively, whereas $\rho$ along the *c*-axis was not reported.

This study investigates the potential of CoSn thin films for interconnect applications.

Demonstrating both the low $\rho$ in the $c$-axis and the large anisotropy of $\rho$ in CoSn thin films is critical for this purpose. We fabricated single-crystalline CoSn ($10\bar{1}0$) films with the $c$-axis in the film plane on an MgO (110) substrate via CoFe/Co/Ru buffer layers. Although the surface of the CoSn ($10\bar{1}0$) was found to be rough due to three-dimensional crystal growth, we confirmed a low $\rho$ of ~13 µΩ cm along the $c$-axis and an approximately tenfold anisotropy of $\rho$ between the $c$-axis and the [$2\bar{1}\bar{1}0$] $a$-axis.

CoSn thin films were co-deposited by magnetron sputtering with Ar gas from Co (purity: 99.9%) and Sn (purity: 99.99%) targets. The chamber base pressure was ~$3\times10^{-6}$ Pa, and the composition of the CoSn films was determined using a combination of inductively coupled plasma optical emission spectroscopy and X-ray fluorescence. We fabricated two types of film structures. One type was polycrystalline CoSn (30 nm) films directly deposited on a thermally oxidized Si substrate, which resulted in randomly oriented polycrystalline films that were convenient for phase identification by X-ray diffraction (XRD).

The other type was single-crystalline CoSn ($10\bar{1}0$) films epitaxially grown on an MgO (110) single-crystalline substrate via $Co_{50}Fe_{50}$/Co/Ru buffer layers. As shown in Fig. 1(b), the ($10\bar{1}0$) plane (M-plane) of CoSn exhibits a relatively small lattice mismatch with that of Ru (the lattice mismatch ratio: –0.5% along [0001] and –2.5% along [$2\bar{1}\bar{1}0$]). Higuchi et al. reported an epitaxial relationship of $(110)_{MgO}[001]_{MgO}$ ∥ $(211)_{Cr}[0\bar{1}1]_{Cr}$ ∥ $(10\bar{1}0)_{Co}[0001]_{Co}$ ∥ $(10\bar{1}0)_{Ru}[0001]_{Ru}$.[28] The hcp-Co layer buffers the lattice mismatch between bcc-Cr (211) and hcp-Ru ($10\bar{1}0$). In our experiment, a bcc-$Co_{50}Fe_{50}$ (hereafter, CoFe) buffer layer was more effective than a Cr buffer layer for growing the Co/Ru layers with higher crystallinity. The MgO substrate was preheated at 600 °C for 10 min in the sputtering chamber to clean its surface and then cooled to RT. MgO (10 nm)/CoFe (2 nm)/Co (2nm)/Ru (2 nm) buffer layers were sputter-deposited at RT, and the CoSn (10–50 nm) films were deposited at $T_{dep}$ = 200–500 °C. The MgO (10 nm) homoepitaxial buffer layer on the MgO (110) substrate improved the reproducibility of the epitaxial growth of the CoFe/Co/Ru/CoSn layers, as described in the supplementary material. We characterized the crystalline structure and microstructure of the CoSn films with a laboratory XRD with Cu-$K_\alpha$ line and scanning transmission electron microscopy (STEM), respectively. The surface morphology of the films was measured by atomic force microscopy (AFM).

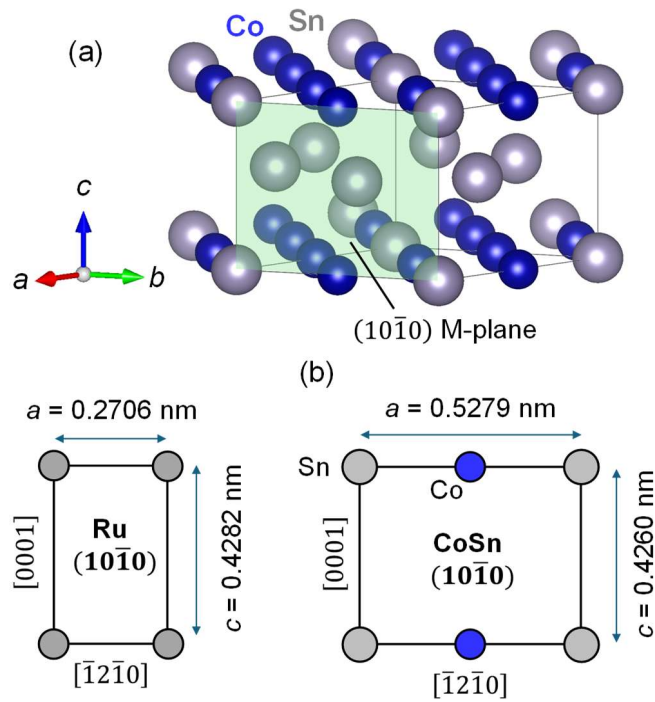

**FIG. 1.** (a) Lattice of CoSn structure and (b) lattice matching between ($10\bar{1}0$) plane (M-plane) of Ru and CoSn. [$\bar{1}2\bar{1}0$] is the Miller-Bravais index for the *b*-axis (equivalent to the *a*-axis, which is expressed as [100] and [$2\bar{1}\bar{1}0$] by the Miller index and Miller-Bravais index, respectively).

Figure 2 shows the out-of-plane XRD profiles of Co-Sn (30 nm) films with different compositions deposited on a thermally oxidized Si substrate at $T_\text{dep}$ = 400 °C. For the $Co_{50.1}Sn_{49.9}$ and $Co_{52.3}Sn_{47.7}$ films, all the diffraction peaks were identified as belonging to the CoSn phase, indicating a single-phase polycrystalline CoSn film with no crystallographic texture. On the other hand, the $Co_{49.6}Sn_{50.4}$ film exhibited diffraction peaks from the $CoSn_2$ phase with $CuAl_2$-type tI12 structure in addition to the CoSn peaks. For the films with higher Sn concentrations, the $CoSn_2$ peaks appeared more distinct. The $Co_{53.5}Sn_{46.5}$ film indicated precipitates of the $Co_3Sn_2$ phase with $Ni_3Sn_2$-type oP20 structure. These results suggest that the composition range for the single-phase CoSn is less than 4 at. %, which is consistent with the line compound nature of the CoSn phase in the equilibrium phase diagram.[29]

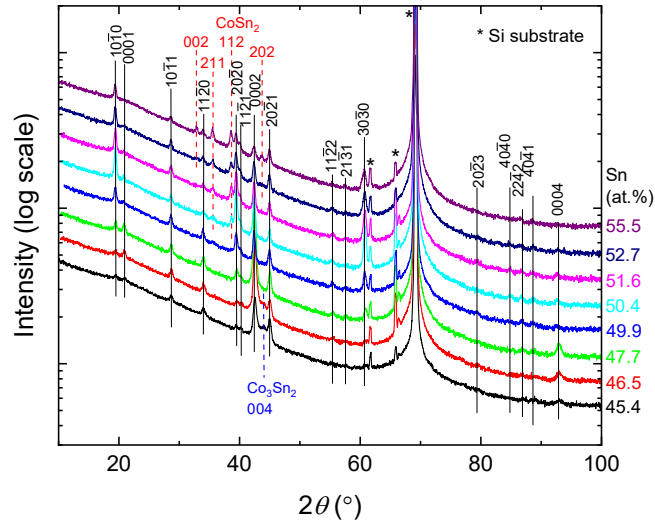

**FIG. 2.** Out-of-plane XRD profiles of Co-Sn (30 nm) films with different compositions directly deposited on a thermally oxidized Si substrate at $T_\text{dep}$ = 400 °C.

Next, we deposited stoichiometric CoSn (30 nm) films on an MgO (110) substrate/MgO (10 nm)/CoFe (2 nm)/Co (2 nm)/Ru (2 nm) buffer structures at $T_\text{dep}$ = 200–500 °C. Figure 3(a) shows the out-of-plane XRD profiles. For $T_\text{dep}$ = 200 and 300 °C, strong peaks from the CoSn ($20\bar{2}3$) plane were observed; hence, CoSn [0001] was not in-plane. On the other hand, the CoSn films deposited at $T_{dep} \geq 350$ °C exhibited $10\bar{1}0$ peak and its higher-order reflections, indicating epitaxial growth of CoSn ($10\bar{1}0$). Figure 3(b) shows the $\phi$-scan profiles for $T_{dep} = 400$ °C, which exhibit the twofold symmetry of the CoSn ($11\bar{2}0$) and ($10\bar{1}1$) planes. The X-ray was irradiated parallel to the [001] direction of the MgO (110) substrate when $\phi = 0°$ and $2\theta = 0°$. This confirms a single-crystalline CoSn film with an orientation relationship of $(110)_{MgO}[001]_{MgO} \parallel (10\bar{1}0)_{CoSn}[0001]_{CoSn}$.

Figures 3(c)-(g) show the AFM images of the surface morphology of the CoSn films deposited at different $T_{dep}$. The sample deposited at $T_{dep} = 200$ °C (Fig. 3(c)) exhibited a relatively flat surface with an arithmetic mean roughness ($R_a$) of 0.15 nm and a peak-to-valley (p-v), the maximum height ($z$)-scale difference, of 3.7 nm. Those deposited at $T_{dep} \geq 350$ °C exhibited significantly increased surface roughness. Additionally, the surface morphology of the CoSn films, particularly those deposited at $T_{dep} = 400$ and 500 °C, exhibits anisotropy; the size of the grain-like morphology is larger along the $a$-axis than the $c$-axis. As the STEM images below show, the large surface roughness is due to the island-like growth of the CoSn film. The typical lateral size of the CoSn islands was 200–300 nm along the $c$-axis for $T_{dep} = 400$ °C, as seen in the AFM image. Sputtered thin films often exhibit large surface roughness when deposited at elevated temperatures due to the high surface mobility of atoms at high temperatures. However, the surface roughness of single-crystalline CoSn thin film strongly depended on the crystal plane of the surface. For comparison, we deposited a single-crystalline CoSn (0001) film at $T_{dep} = 400$ °C on a sapphire (0001) substrate via Pt (3 nm)/Ru (5 nm) buffer layers, as reported by Thapaliya et al.[25] The surface roughness of this film was only $R_a = 0.2$ nm and p-v = 5.5 nm (data not shown here), much smaller than that of the CoSn ($10\bar{1}0$) deposited at $T_{dep} = 400$ °C ($R_a = 1.8$ nm and p-v = 25 nm). This large difference in surface roughness between these crystal planes may be due to differences in their surface energies.

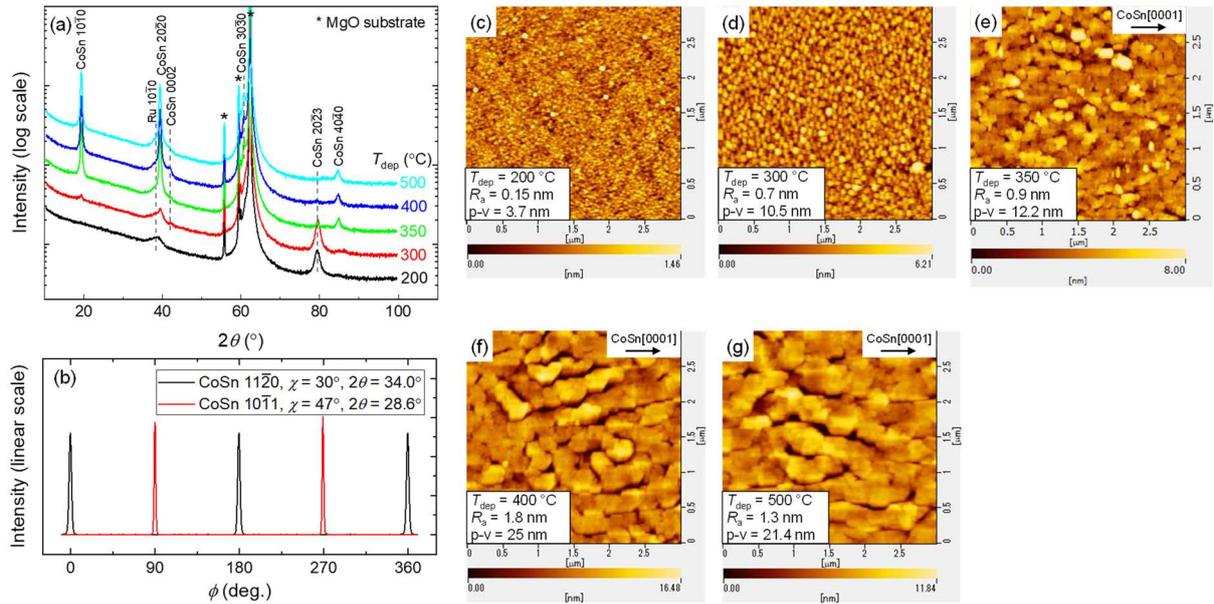

**FIG. 3.** (a) Out-of-plane X-ray diffraction (XRD) profiles of MgO (110) substrate/MgO (10 nm)/CoFe (2 nm)/Co (2 nm)/Ru (2 nm)/CoSn (30 nm) samples with different $T_{dep}$ for the CoSn layer, and (b) $\phi$-scan from CoSn (11$\bar{2}$0) and (10$\bar{1}$1) planes for $T_{dep}$ = 400 °C. (c)-(g) Surface roughness images of the CoSn (30 nm) film deposited at $T_{dep}$ = 200–500 °C, respectively, by AFM.

The microstructure of the 30-nm-thick single-crystalline CoSn film deposited at $T_{dep}$ = 400 °C was analyzed using STEM. Figure 4(a) shows a low-magnification high-angle annular dark-field (HAADF)-STEM image taken from the [0001] zone axis of CoSn. The CoSn layer consists of trapezoidal islands ranging in thickness from 28 to 44 nm, consistent with the large p-v value of 25 nm observed in the AFM image (Fig. 3(f)).

Fig. 4(b) shows a magnified HAADF-STEM image and the corresponding energy dispersive X-ray spectroscopy (EDS) elemental map of Pt as a protective coating (blue), Sn (green), Ru (purple), Fe (yellow), and O (red). Separate EDS maps of all relevant elements are provided in Fig. S6(a) in the supplementary material. The EDS elemental map and compositional line profiles across the constituent layers reveal a uniform distribution of Co and Sn throughout the CoSn layer. The top surface was covered with a thin, 1.5-nm-thick oxidized layer. The CoSn layer is epitaxially grown with $[10\bar{1}0]$ orientation on the Ru buffer layer. The orientation relationship, determined from nanobeam electron diffraction (NBED) patterns (see Fig. S2 in the supplementary material), is described as $(110)_{MgO}[001]_{MgO}$ ∥ $(211)_{CoFe}[0\bar{1}1]_{CoFe}$ ∥ $(10\bar{1}0)_{Co}[0001]_{Co}$ ∥ $(10\bar{1}0)_{Ru}[0001]_{Ru}$ ∥ $(10\bar{1}0)_{CoSn}[0001]_{CoSn}$.

Figure 4(c) shows a magnified HAADF-STEM image of the faceted CoSn surface. The NBED pattern of the CoSn layer confirms that the facets correspond to the $(10\bar{1}0)$, $(11\bar{2}0)$, and $(01\bar{1}0)$ planes. Faint diffraction contrast is also observed along the $\{10\bar{1}0\}$ trace, as indicated by arrows, which indicates the presence of domain boundaries. These boundaries are more clearly seen in the enlarged image in Fig. 4(c), where the kagome lattice is resolved: bright Sn columns and dim Co columns are clearly visible, with a local shift in the kagome lattice across the boundary. Such domain boundaries are likely formed by multiple nucleation events followed by coalescence of CoSn grains.

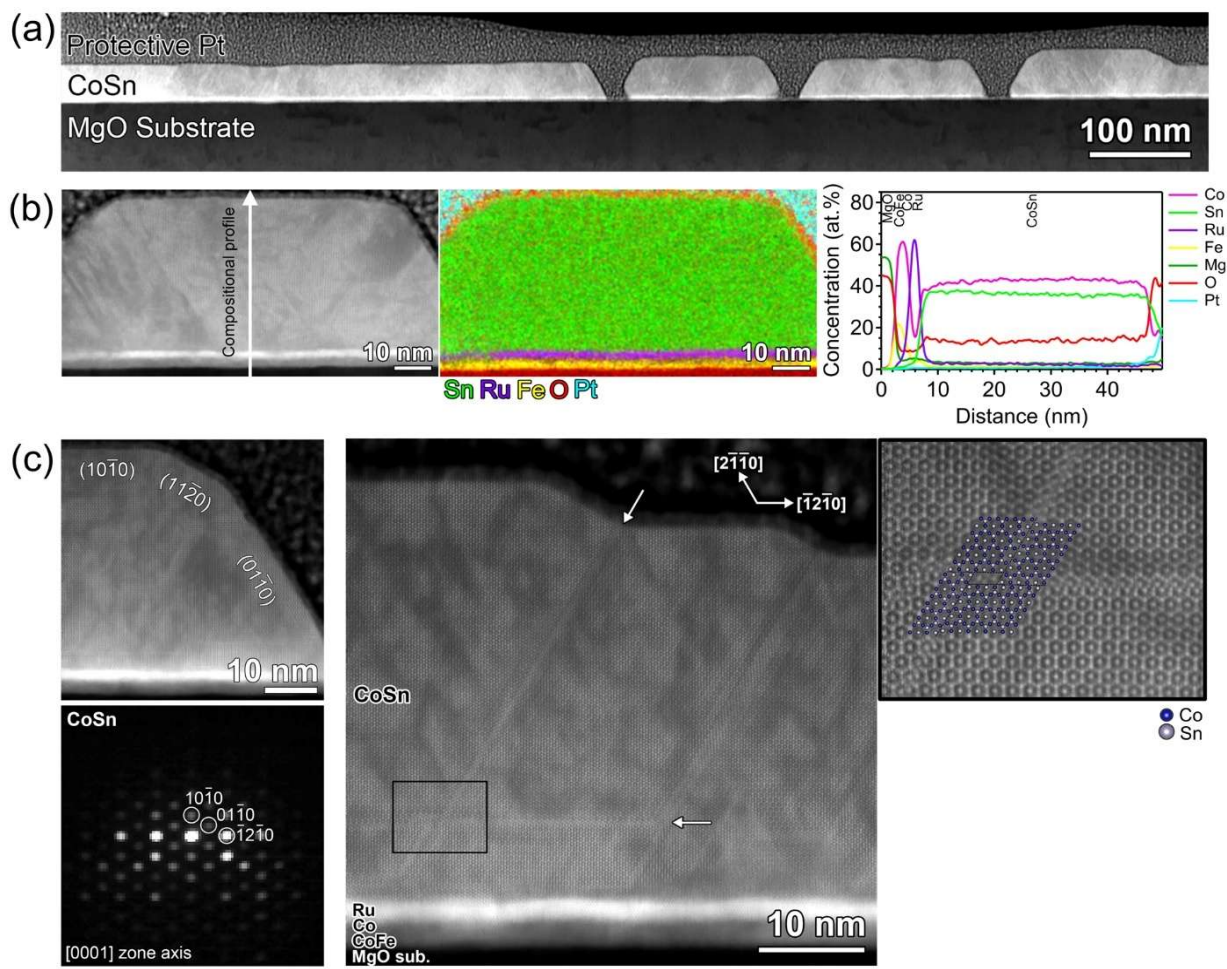

**FIG. 4.** HAADF-STEM images of the MgO (110) substrate/CoFe (2 nm)/Co (2 nm)/Ru (2 nm)/CoSn (30 nm) [$T_{dep}$ = 400 °C] film viewed along the CoSn [0001] direction and taken from various regions. (a) Low-magnification image showing CoSn morphology. (b) Magnified HAADF-STEM image and its corresponding EDS elemental map and line compositional profile. The Pt layer was deposited as a protective coating during the specimen preparation. (c) Magnified HAADF-STEM images (with enlarged image taken from a rectangular box), and the NBED taken from CoSn layer.

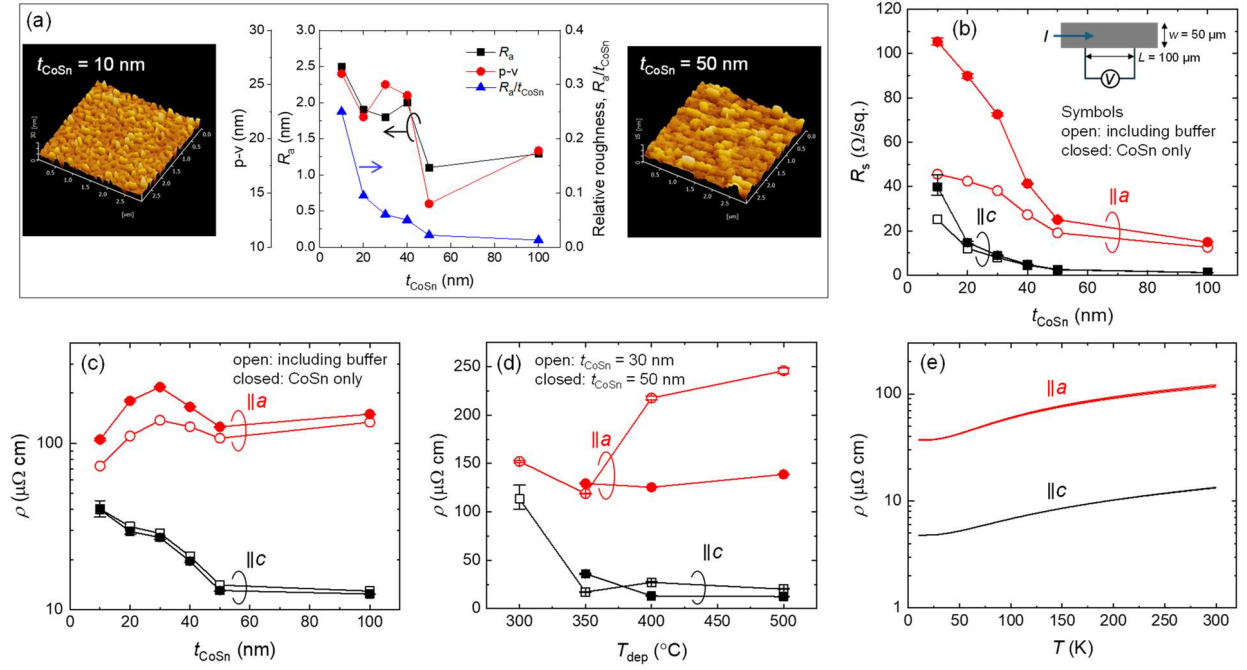

**FIG. 5.** Thickness ($t_{CoSn}$) dependence of (a) the surface roughness, (b) sheet resistance ($R_s$), and (c) resistivity of the single-crystalline CoSn films at $T_{dep}$ = 400 °C. In (b) and (c), the open symbols are the average $R_s$ and $\rho$ including the buffer layer, and the closed symbols are the $R_s$ and $\rho$ of only the CoSn layer. (d) $T_{dep}$-dependence of $\rho$ of the CoSn films [$t_{CoSn}$ = 30 nm (open symbols) and 50 nm (closed symbols)]. (e) Temperature dependence of the resistivity of CoSn (50 nm) at $T_{dep}$ = 400 °C

We evaluated the resistivity of single-crystalline CoSn ($10\bar{1}0$) films deposited on the MgO (10 nm)/CoFe (2 nm)/Co (2 nm)/Ru (2 nm) buffer layers. Figure 5(a) shows the $t_{CoSn}$-dependence of the surface roughness of the CoSn films deposited at $T_{dep}$ = 400 °C. For $t_{CoSn}$ = 10 nm, the p-v value was 26 nm, indicating an island growth of CoSn, as seen in the STEM image of the $t_{CoSn}$ = 30 nm sample (Fig. 4(a)). The p-v values remained large at ~25 nm for $t_{CoSn}$ = 10–40 nm. However, $t_{CoSn}$ = 50 nm exhibited a significantly reduced p-v value of 14 nm, suggesting that the valley of the surface morphology was partially filled. As shown in Fig. 5(a), the relative roughness to thickness ($R_a/t_{CoSn}$) increases as $t_{CoSn}$ decreases.

The sheet resistance ($R_s$) of the CoFe (2 nm)/Co (2 nm)/Ru (2 nm)/CoSn ($t_{\text{CoSn}}$) films were measured in strip-shaped devices patterned into a width ($w$) of 50 μm and a length ($L$) of 100 μm (see the inset of Fig. 5(b)) as $R_s = R\frac{w}{L}$, where $R$ is the measured resistance. Note that the resistance measurements of unpatterned films using an in-line four-probe yielded inaccurate $R_s$ and resistivity values for the CoSn films with resistivity anisotropy. See the supplementary material for details. The open symbols in Fig. 5(b) shows the $R_s$ including the CoFe/Co/Ru buffer layers at RT along the *c*-axis and *a*-axis [$T_{\text{dep}}$ = 400 °C for CoSn]. The values of $R_s$ along the *c*-axis were lower than those along the *a*-axis, indicating the anisotropic resistivity of CoSn. The open symbols in Fig. 5(c) shows the average resistivity ($\rho^{\text{ave}}$) including the buffer layers. The film with $t_{\text{CoSn}}$ = 50 nm showed $\rho_{\|c}^{\text{ave}}$ = 14 μΩ cm and $\rho_{\|a}^{\text{ave}}$ = 107 μΩ cm. To evaluate the resistivity of the CoSn films only, we subtracted the $R_s$ of the CoFe/Co/Ru buffer layers ($R_s$ = $68.1_{-10.5}^{+15.2}$ and 80.5 ± 0.9 Ω/sq. along the *c*-axis and *a*-axis, respectively) estimated by separate experiments as described in the supplementary material.

By subtracting the $R_s$ values of the buffer layers, the resistivity of the CoSn film was obtained to be $\rho_{\|c}^{\text{CoSn}} = 39.9_{-3.8}^{+5.3}$ μΩ cm and $\rho_{\|a}^{\text{CoSn}} = 105.3_{-1.5}^{+1.6}$ μΩ cm for $t_{\text{CoSn}}$ = 10 nm, and $\rho_{\|c}^{\text{CoSn}}$ = 13.0±0.1 μΩ cm and $\rho_{\|a}^{\text{CoSn}}$ = 125.0±0.4 μΩ cm for $t_{\text{CoSn}}$ = 50 nm, as shown by the closed symbols in Fig. 5(c). These results demonstrate a significant anisotropy of resistivity in the single-crystalline CoSn films. While the value of $\rho_{\|a}^{\text{CoSn}}$ for $t_{\text{CoSn}}$ = 50 nm was close to those reported to the bulk sample ($\rho_{\|a}^{\text{CoSn}}$ = 120 μΩ cm),[22,24] the $\rho_{\|c}^{\text{CoSn}}$ value for $t_{\text{CoSn}}$ = 50 nm was larger than those of the bulk sample ($\rho_{\|c}^{\text{CoSn}}$ = 3–7 μΩ cm).[20–24] As shown in Fig. 3(f), the surface morphology of the CoSn film was also anisotropic; the width of the CoSn islands was larger along the *c*-axis than the *a*-axis, which may contribute to the measured resistivity anisotropy of the CoSn films. However, as shown in the supplementary material, another CoSn film with a

different type of surface morphology, where the CoSn islands were broader along the *a*-axis than the *c*-axis, exhibited similar resistivity anisotropy of $\rho_{\parallel c}^{\text{CoSn}} < \rho_{\parallel a}^{\text{CoSn}}$. Therefore, the observed resistivity anisotropy is considered to reflect the intrinsic resistivity anisotropy of CoSn as reported in the bulk single crystals. [20–24]

The thickness dependence of resistivity is critical for the interconnect applications. The $\rho_{\parallel c}^{\text{CoSn}}$ value of the single-crystalline CoSn films clearly depended on $t_{\text{CoSn}}$ below 50 nm, i.e., $\rho_{\parallel c}^{\text{CoSn}}$ increased with decreasing $t_{\text{CoSn}}$, as shown in Fig. 5(c). Since the CoSn ($10\bar{1}0$) single-crystalline films exhibited three-dimensional island-like growth and the relative roughness ($R_a/t_{\text{CoSn}}$) increased with decreasing $t_{\text{CoSn}}$, as depicted in Fig. 5(a), the increase in $\rho_{\parallel c}^{\text{CoSn}}$ with decreasing $t_{\text{CoSn}}$ may include a contribution from film roughness. Therefore, it is not possible to discuss the intrinsic thickness dependence of the resistivity of the present CoSn ($10\bar{1}0$) films. The realization of smoother CoSn single-crystalline films with the *c*-axis in-plane is highly desired.

Figure 5(d) shows the dependence of $\rho_{\parallel c}^{\text{CoSn}}$ and $\rho_{\parallel a}^{\text{CoSn}}$ on $T_{\text{dep}}$ for $t_{\text{CoSn}}$ = 30 and 50 nm. At $T_{\text{dep}}$ = 300 °C for $t_{\text{CoSn}}$ = 30 nm, the CoSn ($20\bar{2}3$) was parallel to the film plane, and the CoSn [0001] was not in-plane [Fig. 3(a)], therefore, the difference in resistivity between the two orientations was relatively small compared to the cases with $T_{\text{dep}} \geq 350$ °C, at which the CoSn ($10\bar{1}0$) epitaxially grew on Ru ($10\bar{1}0$). At $T_{\text{dep}}$ = 400 and 500 °C, both $\rho_{\parallel c}^{\text{CoSn}}$ and $\rho_{\parallel a}^{\text{CoSn}}$ for $t_{\text{CoSn}}$ = 30 nm were much higher than those for $t_{\text{CoSn}}$ = 50 nm. This could be due to the reduced relative roughness of the $t_{\text{CoSn}}$ = 50 nm films compared to the $t_{\text{CoSn}}$ = 30 nm. For $t_{\text{CoSn}}$ = 50 nm, little change occurred in $\rho_{\parallel c}^{\text{CoSn}}$ between $T_{\text{dep}}$ = 500 °C (12.4 μΩ cm) and $T_{\text{dep}}$ = 400 °C (13.0 μΩ cm).

Figure 5(e) shows the temperature dependence of $\rho_{\parallel c}^{\text{CoSn}}$ and $\rho_{\parallel a}^{\text{CoSn}}$ for $t_{\text{CoSn}}$ = 50 nm and $T_{\text{dep}}$ = 400 °C. Both $\rho_{\parallel c}^{\text{CoSn}}$ and $\rho_{\parallel a}^{\text{CoSn}}$ showed monotonic decreases with decreasing $T$, consistent with the bulk single crystal.[24] However, the residual resistivity of this thin film was much larger than that of the bulk single crystal: $\rho_{\parallel c}^{\text{CoSn}}$ = 4.8 μΩ cm and $\rho_{\parallel a}^{\text{CoSn}}$ = 37.1 μΩ cm at 10 K for the thin film, and $\rho_{\parallel c}^{\text{CoSn}}$ = 0.19 μΩ cm and $\rho_{\parallel a}^{\text{CoSn}}$ = 11.44 μΩ cm at 2 K for the bulk single crystal.[24] These results suggest temperature-independent scattering sources for conduction electrons in the thin films, such as surface roughness, impurities, and crystal defects. Identifying the cause of the higher $\rho$ in the present thin-film CoSn is critical for further reducing $\rho$.

In conclusion, single-phase CoSn films were deposited by sputtering on heated substrates at ~400 °C. Using bcc-CoFe/hcp-Co/hcp-Ru buffer layers on an MgO (110) substrate, epitaxial growth of CoSn (10$\bar{1}$0) single-crystalline films was achieved. The films exhibited significant surface roughness arising from three-dimensional growth, particularly in thinner films, and STEM observations revealed the presence of domain boundaries. Despite these structural imperfections, the CoSn films showed low resistivity along the *c*-axis, reaching 13 μΩ cm, and higher resistivity along the orthogonal *a*-axis (>100 μΩ cm), consistent with the anisotropic resistivity reported for bulk single crystals. Further improvements in surface morphology will be crucial for accurately assessing the intrinsic thickness dependence of resistivity in CoSn thin films.

**Acknowledgements**

We are grateful to Toyohiro Chikyow, Toshihide Nabatame, Hirofumi Suto, Yuya Sakuraba, Rohit Dahule, and Ryoji Sahara (NIMS), and Mitsuru Ohtake (Yokohama National University) for valuable discussions and suggestions, and Takanobu Hiroto and Murali Krishnan Manikketh (NIMS) for experimental supports. This research was supported in part by KIOXIA Corporation and ARIM of MEXT (JPMXP1225NM5220).

**Supplementary material**

Refer to the supplementary material for the sample fabrication process, the epitaxial relationship, the sheet resistance measurements, the effect of surface morphology anisotropy on resistivity anisotropy, and additional EDS and STEM data.

**Data Availability Statement**

The data supporting the findings of this study are available from the corresponding author upon reasonable request.

**Conflict of interest**

The authors have no conflict of interest regarding the publication of this article.

Supplementary Materials

**Fabrication and characterization of single-crystalline CoSn ($10\bar{1}0$) kagome metal thin films for interconnect applications: structure and anisotropic electrical resistivity**

Tomoya Nakatani,[1*] Nattamon Suwannaharn,[2] and Taisuke T. Sasaki[2]

[1]Research Center for Magnetic and Spintronic Materials, National Institute for Materials Science, 1-2-1, Senen, Tsukuba, Ibaraki 305-0047, Japan.

[2]Research Center for Structural Materials, National Institute for Materials Science, 1-2-1, Senen, Tsukuba, Ibaraki 305-0047, Japan.

[*]E-mail: nakatani.tomoya@nims.go.jp

1. **Sample preparation**

MgO(110) substrates were cleaned sequentially with acetone, deionized water, and isopropanol using ultrasonication. Prior to film deposition, the substrates were heated at 600 °C for 10 min in a sputtering chamber with a base pressure of $\sim 3\times 10^{-6}$ Pa, which is believed to remove the magnesium hydroxide from the MgO surface and enable epitaxial film growth. No cleaning or heating process was carried out for thermally oxidized Si substrates.

CoSn films were co-sputtered from Co (purity: 99.9%) and Sn (purity: 99.99%) targets using RF power. The choice of RF power was merely due to the configuration of our sputtering tool, which consists of three deposition chambers manufactured by Ulvac, Inc. One chamber has eight 2-inch cathodes with DC power supplies but no substrate heater. Another chamber has a 4-inch cathode with an RF power supply, which we used for depositing the MgO homoepitaxial buffer layer. Third chamber has has two 2-inch cathodes with RF power supplies and a substrate heater. We used this chamber to deposit CoSn. We believe that DC power would also work for co-sputtering CoSn. The target-substrate distance was approximately 20 cm. RF powers of 100 W on the Co target and ~50 W on the Sn targets produced nearly stoichiometric CoSn (50:50 at. %) films at a deposition rate of ~0.06 nm/s. 10 sccm of Ar gas was flown from each Co and Sn cathode, and the chamber pressure was approximately 0.2 Pa.

As discussed in the main text of this paper, controlling the composition is essential for producing single-phase CoSn films. We used a combination of inductively coupled plasma optical emission spectroscopy (ICP-OES) and wavelength-dispersive X-ray fluorescence (XRF) to analyze the CoSn film composition. Although ICP-OES is considered to be the most accurate method, it requires sufficient sample mass (typically ~1 mg) and careful operation by an expert. Conversely, XRF measurements are convenient, quick, and precise when recipes are created based on standard samples with accurate composition and thickness data. First, we deposited ~200-nm-thick CoSn films with three different compositions on a bare Si substrate and analyzed their compositions with ICP-OES. Next, we deposited ~30-nm-thick CoSn films with the same deposition conditions as those for the ICP-OES analysis. The thickness of the films was measured by X-ray reflectivity. Using these samples as standards, we created an XRF measurement recipe. We typically use CoSn (20–30 nm) films for XRF analysis.

To evaluate the precision of the XRF analysis, we measured a CoSn (20 nm) film deposited on a thermally oxidized Si substrate thirteen times. The average Sn composition was 50.10 at. % with a standard deviation of 0.09 at. %. These results indicate that the XRF analysis has sufficiently high precision. However, the "accuracy" of the composition analysis relies on the accuracy of the ICP-OES measurements, which we are unaware of.

Figures S1 shows the out-of-plane XRD profiles of stoichiometric CoSn (~30 nm) films deposited on a thermally oxidized Si substrate at room temperature (RT). (a) and (b) show the profiles of the samples without and with a post-deposition-annealing at $T_{ann}$ = 400 °C for 30 min, repsectively. The as-deposited CoSn film at RT (Fig. S1(a)) exhibited two weak peaks. The peak at $2\theta$ = 42.6° corresponds to the CoSn 0002 reflection. Annealing the CoSn film produced peaks from the CoSn (0001) plane and its higher-order reflections, indicating a [0001]-out-of-plane texture. Figure S1(c) shows the data for the CoSn film deposited at $T_{dep}$ = 400 °C. All the peaks were identified as belonging to the CoSn phase, indicating a polycrystalline single-phase CoSn film with no particular crystallographic texture. These three samples were deposited under the identical deposition power and Ar gas flow conditions for the same amount time. The thickness and Sn concentration of these films, as measured by XRF, were (a) 29.2 nm and 50.5 at. %, (b) 29.1 nm and 50.1 at. %, and (c) 29.3 nm and 50.1 at. %, respectively. Therefore, the deposition rate and film composition of the CoSn films were approximately constant under these $T_{dep}$ and $T_{ann}$ conditions. We found that the CoSn composition and deposition rate were nearly constant for $T_{ann}$ between RT and 500 °C.

Table S1 shows the deposition conditions for the CoSn($10\bar{1}0$) single-crystalline samples shown in Fig. 3(a). The MgO heteroepitaxial buffer layer was deposited by RF sputtering a sintered MgO target at RT. The importance of the MgO buffer layer is explained in Section 4 of this document. Due to the aforementioned limitation of our sputtering tool, we deposited the CoFe/Co/Ru buffer layers only at RT.

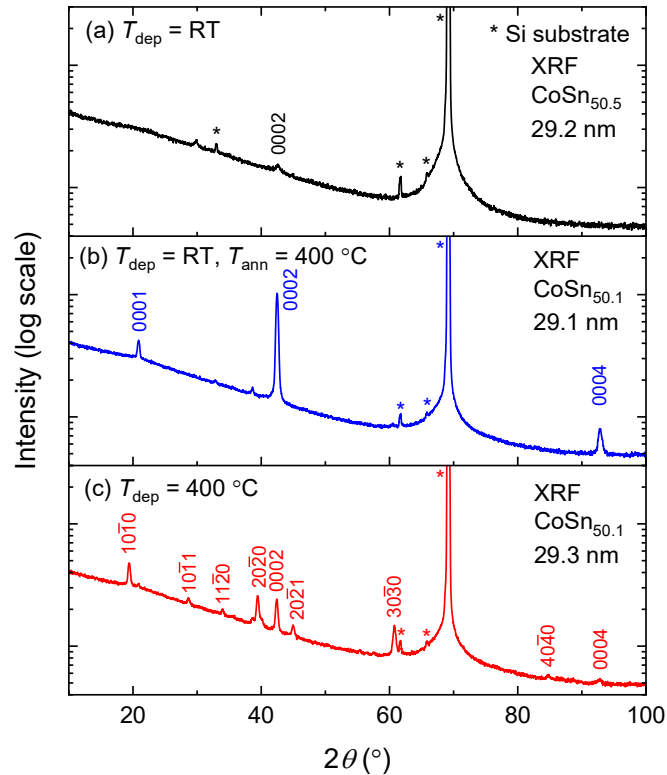

**FIG. S1.** Out-of-plane XRD profiles of stoichiometric CoSn (~30 nm) films deposited on a thermally oxidized Si substrate. (a) Deposited at RT, (b) deposited at RT and annealed at $T_{ann}$ = 400 °C, and (c) deposited at $T_{dep}$ = 400 °C.

**TABLE S1.** Deposition conditions for MgO(110) substrate/MgO (10 nm)/CoFe (2 nm)/Co (2 nm)/Ru (2 nm)/CoSn (30 nm) single-crystalline films.

| Material | MgO | Co$_{50}$Fe$_{50}$ | Co | Ru | CoSn |
|---|---|---|---|---|---|
| Deposition power (W) | RF 200 | DC 50 | DC 50 | DC 50 | RF 100 (Co) /RF 48 (Sn) |
| Ar flow rate /pressure | 70 sccm /0.8 Pa | 40 sccm /0.2 Pa | 30 sccm /0.15 Pa | 30 sccm /0.15 Pa | 10+10 sccm /0.2 Pa |
| Substrate temperature (°C) | RT | RT | RT | RT | 350–500 |
| Deposition rate (nm/s) | 0.022 | 0.021 | 0.024 | 0.043 | 0.061 |

## 2. Epitaxial relationship in MgO(110)/CoFe/Co/Ru/CoSn layers

The samples discussed in our paper have the following structure: MgO(110) substrate/MgO (10 nm) homoepitaxial buffer/CoFe (2 nm)/Co (2 nm)/Ru (2 nm)/CoSn (10-100 nm). The CoFe/Co/Ru trilayers function as heteroepitaxial buffer layers for the growth of single-crystalline CoSn films, as confirmed with XRD $\phi$-scans (Fig. 3(b)) and STEM observations (Fig. 4). However, due to the 2-nm thickness of the CoFe, Co, and Ru buffer layers, we could not identify the orientation relationship between these layers. Therefore, we analyzed the orientation relationship in a sample with thicker buffer layers: MgO(110) substrate/CoFe (5 nm)/Co (5 nm)/Ru (10 nm)/CoSn (30 nm), with a deposition temperature of CoSn ($T_{\text{dep}}$) of 400 °C.

Nanobeam electron diffraction (NBED) patterns of each layer are shown in Fig. S2. NBED confirmed the following orientation relationships from bottom to top: MgO(110)[001] ∥ CoFe(211)[0$\bar{1}$1] ∥ Co(10$\bar{1}$0)[0001] ∥ Ru(10$\bar{1}$0)[0001] ∥ CoSn(10$\bar{1}$0)[0001], consistent with that reported for the MgO(110)/Cr/Co/Ru epitaxy. [1] A schematic crystal illustration was created to demonstrate this relationship, as shown in Fig. S2. Additionally, two sets of diffraction spots were observed in the CoFe layer, mirrored with respect to the (211) plane, as indicated by the white solid rectangle and blue dashed rectangles. This observation suggests a Σ3 twin boundary in the CoFe layer, where the orientation difference between the parent and twinned domains is a 60° rotation around the [1̄11] axis. Note that the extra spots are double diffraction, which arises when the electron beam sequentially diffracted by adjacent domains. Furthermore, the epitaxial growth of the overlying Co buffer layer showed a (101̄0) plane, regardless of the twin domain, since both twin domains owned the same (211) plane.

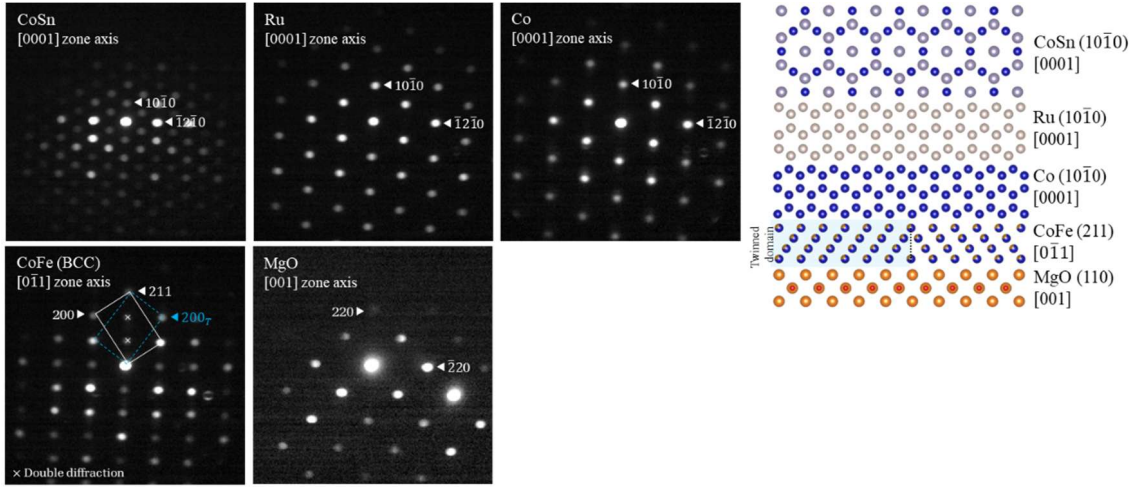

**FIG. S2**. Nanobeam electron diffraction patterns acquired from an MgO(110) substrate/CoFe (5 nm)/Co (5 nm)/Ru (10 nm)/CoSn (30 nm) with $T_{dep}$ = 400 °C, and a corresponding schematic crystal illustration.

## 3. Sheet resistance measurement

The sheet resistance of thin films is often measured by placing an in-line four-probe on the sample without patterning, as illustrated in Fig. S3(a). For films whose lateral sizes ($l$) are much larger than the probe pitch ($s$), the sheet resistance ($R_s$) of the film is given by

$$R_s = \frac{\pi}{\ln 2} \frac{V}{I}, \tag{S1}$$

where $I$ is the bias current applied between probes 1 and 4, and $V$ is the voltage measured between probes 2 and 3. When $l/s$ is approximately less than 40, a geometrical correction factor must be added to Eq. S1, as explained in Ref. [2, 3]

However, we found that the sheet resistance measurements in unpatterned films using in-line four-probe yield incorrect $R_s$ values for the CoSn(101̄0) single-crystalline films. Table S2 shows the $R_s$ values of the CoFe (2 nm)/Co (2 nm)/Ru (2 nm)/CoSn (30 nm) [$T_{dep}$ = 400 °C] film deposited on MgO(110) substrate via an MgO (10 nm) homoepitaxial buffer layer. The dimensions of the patterned device were a line width of $w$ = 50 μm and a distance between the voltage probes of $L$ = 100 μm, as shown in Fig. S3(b). While the $R_s$ values measured by an in-line four-probe showed ~40% anisotropy between $R_{s\|c}$ and $R_{s\|a}$, those measured in the patterned device showed a much greater anisotropy of $R_{s\|a}/R_{s\|a}$~4.8. We confirmed that the $R_{s\|c}$ and $R_{s\|a}$ values were consistent for devices with different values of $w$ of 10 and 20 μm. These results indicate that the $R_s$ measurements with an in-line four-probe on unpatterned CoSn films lead to incorrect results due to different current distributions within the CoSn films for currents parallel to the $c$-axis and $a$-axis.

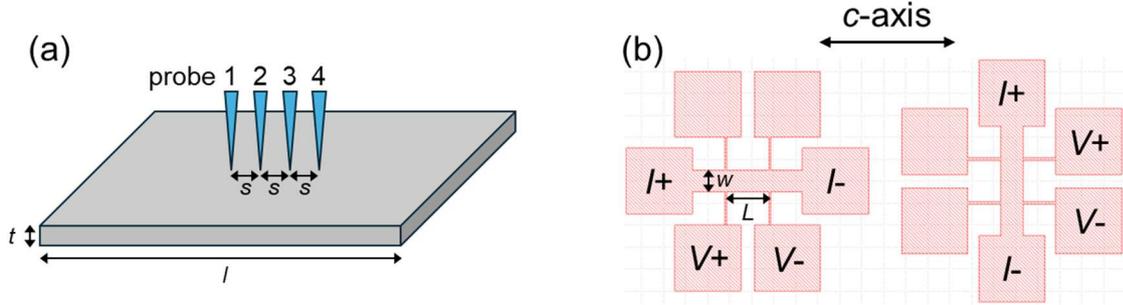

**FIG. S3.** (a) Schematic of the sheet resistance measurement in an unpatterned film using an in-plane four-probe. (b) Design of patterned devices for sheet resistance measurement for CoSn single-crystalline films with resistivity anisotropy.

**TABLE S2.** Sheet resistance ($R_s$) values measured for an unpattered film with in-line four-probe [Fig. S3(a)] and for a patterned device [Fig. S3(b)].

| Orientation | $R_s$ (Ω/sq.) | |
|---|---|---|
| | Unpatterned film with in-line four-probe | Patterned device |
| ∥ $c$ | 20.5 | 8.0 |
| ∥ $a$ | 28.4 | 38.2 |

## 4. Sheet resistance measurement of CoFe/Co/Ru buffer layers

The epitaxial growth of CoSn($10\bar{1}0$) requires the CoFe/Co/Ru buffer layers an MgO(110) substrate. Here, we explain how we estimated the sheet resistance ($R_s$) value of the buffer layers. The $R_s$ of the buffer layers can be easily evaluated by measuring the $R_s$ of the MgO(110) substrate/CoFe/Co/Ru (without a capping layer) or the MgO(110) substrate/CoFe/Co/Ru/insulating capping layer. However, in such structures, the scattering of conduction electrons at the Ru surface and the Ru/insulator interface adds extra resistance, leading to an overestimation of the $R_s$ value of the CoFe/Co/Ru buffer layers. Note that in the MgO(110) substrate/CoFe/Co/Ru/CoSn samples, the interfacial resistance at the epitaxial Ru/CoSn interface is expected to be small.

To address this, we evaluated the $R_s$ of the CoFe/Co/Ru buffer layers in the sample structure of MgO(110) substate/MgO (10 nm)/CoFe (2 nm)/Co (2 nm)/Ru (2 nm)/Ag$_{90}$Sn$_{10}$ ($t_{AgSn}$ = 10–30 nm). We chose an fcc-Ag$_{90}$Sn$_{10}$ (hereafter, AgSn) layer for epitaxial growth on Ru($10\bar{1}0$) and also due to the relatively large $\rho$ of 30–50 nm of AgSn,[3] which ensures a more precise evaluation of the $R_s$ of the CoFe/Co/Ru buffer layers than with low-resistivity Ag. Figure S4(a) shows the out-of-plane XRD profiles of the samples. The Ru $10\bar{1}0$ and AgSn 220 peaks suggest the epitaxial growth of AgSn(110) on Ru($10\bar{1}0$).

In this study, we used two setups for resistance measurement. For the RT measurements shown in Figs. 5(b)–(d), we used a Keithley 2400 source measure unit (SMU) with a constant current of 1.0 mA and a manual prober with tungsten probes that were pressed onto the device [Fig.S3(b)]. We performed the resistance measurements at different temperatures [Fig. 5(e)] with a Quantum Design Dynacool system and its built-in resistance meter. For these measurements, we bonded the devices with Al wires. Device resistance was measured under a constant current of 0.1 mA. The lower current, compared to that used for the RT measurement, was used to avoid heating the device. Due to some measurement issues, the device resistance values obtained by the manual prober and SMU at RT and by Dynacool at 300 K were not identical. For the RT measurements, we subtracted the $R_s$ values of the buffer layers obtained by the manual prober and SMU. For the temperature-variable measurements, we subtracted the values by Dynacool.

Figure S4(b) shows the 1/$R_s$ vs. $t_{AgSn}$ plot for the RT measurements. The inverse of the extrapolated intercept yields the $R_s$ of the CoFe/Co/Ru buffer layer: $R_s = 68.1^{+15.2}_{-10.5}$ Ω/sq. along the $c$-axis of Ru, and $R_s = 80.5 \pm 0.9$ Ω/sq. along the $a$-axis of Ru. The error stems from the error in the linear fitting of the $1/R_s$ vs. $t_{AgSn}$ plot. Figure S4(c) shows the $1/R_s$ vs. $t_{AgSn}$ plot at temperatures ranging from 10 K to 300 K using Dynacool. Even at nearly the same temperature (RT with the manual prober and 300 K with Dynacool), the $R_s$ values measured by these two systems differed: $R_s$ = 21.06, 12.63, and 8.73 Ω/sq. for the manual prober, and $R_s$ = 22.89, 12.39, and 8.98 Ω/sq. for Dynacool, for $t_{AgSn}$ = 10, 20, and 30 nm, respectively, along the $c$-axis. These differences resulted in different $R_s$ values and error ranges: $R_s = 68.1^{+15.2}_{-10.5}$ Ω/sq. and $91.8^{+5.7}_{-24.5}$ Ω/sq. for the manual prober and Dynacool, respectively. Figure S4(d) shows the temperature dependence of $R_s$ of the buffer layers.

The $R_s$ values and their errors of the CoFe/Co/Ru buffer layers were used to calculate the $R_s$ and $\rho$ values of the CoSn films shown in Figs. 5(b)–(e). Despite the large errors of $R_s$ of the buffer layers, the error in $\rho^{CoSn}$ was quite small, especially for thick $t_{CoSn}$ along the low-resistive $c$-axis. However, accurately evaluating $\rho$ in thin film ($t$ ~10 nm) is critical for interconnect applications. Therefore, precisely estimating the $R_s$ value of buffer layers is essential for future studies.

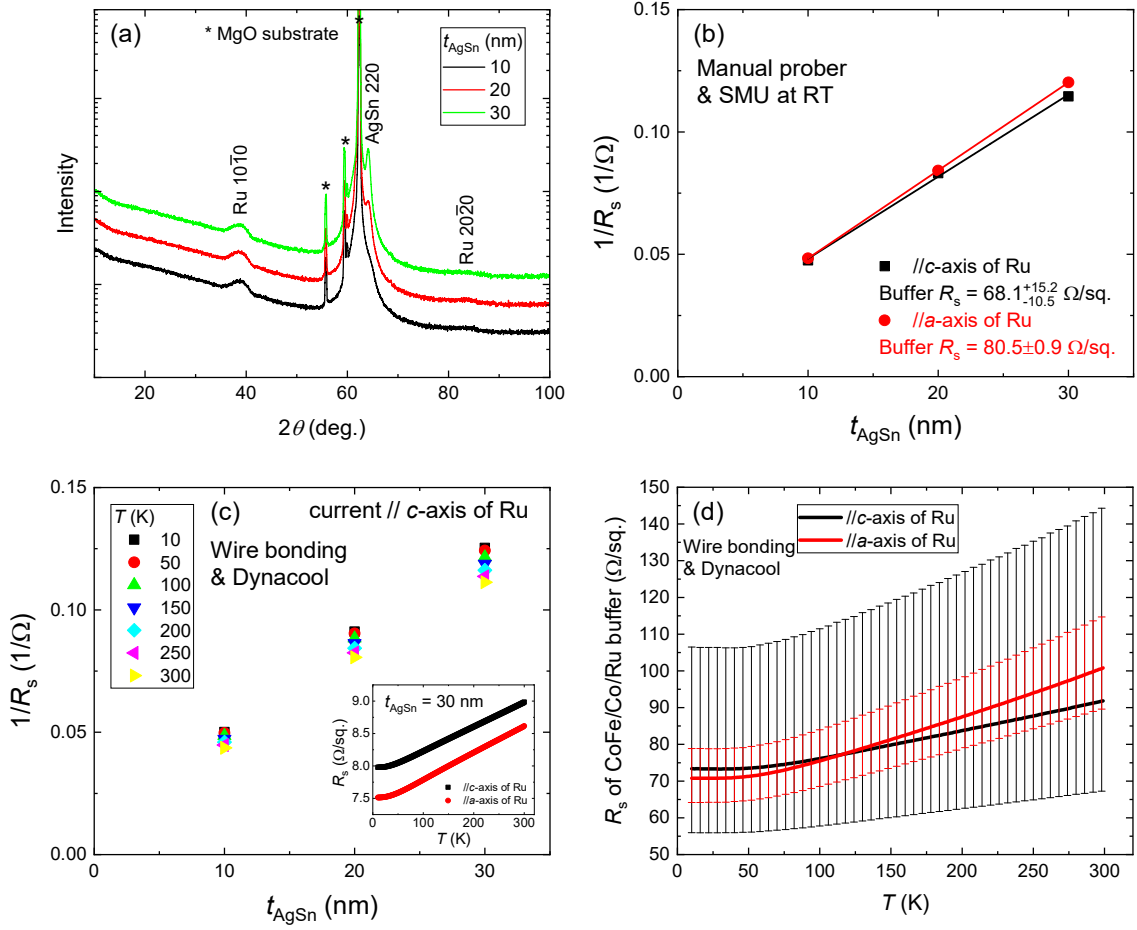

**FIG. S4.** Evaluation of the sheet resistance MgO(110) substrate/CoFe (2 nm)/Co (2 nm)/Ru (2 nm)/AgSn ($t_{AgSn}$) samples. (a) Out-of-plane XRD profiles. $1/R_s$ vs. $t_{AgSn}$ plots at (b) RT and (c) various temperatures. (d) temperature dependence of $R_s$ of the CoFe/Co/Ru buffer layers obtained from the $1/R_s$ vs. $t_{AgSn}$ plots in (c).

## 5. Resistivity anisotropy: the effect of surface morphology anisotropy

In this paper, we demonstrated the resistivity anisotropy between the *c*-axis and *a*-axis of the CoSn thin films. The CoSn single-crystalline films exhibited significant surface roughness due to three-dimensional crystal growth. Notably, the surface morphology also exhibited anisotropy between the *c*- and *a*-axis directions, i.e., the lateral size of the CoSn crystal islands was greater along the *c*-axis than the *a*-axis, as shown in Figs. 3(e)–(g). This surface morphology anisotropy can contribute to anisotropy of sheet resistance.

To determine whether the resistivity anisotropy observed in the CoSn thin films (Fig. 5) is solely due to the surface morphology anisotropy, we examined data from a different CoSn sample with a different type of surface morphology. Figure S5(a) shows the out-of-plane XRD pattern of an MgO(110) substrate/CoFe (2 nm)/Co (2 nm)/Ru (2 nm)/CoSn (30 nm) [$T_{\rm dep}$ = 400 °C] sample without an MgO (10 nm) homoepitaxial buffer layer. Then, we experienced a serious reproduction issue with the CoSn film. Sometimes, we sometimes obtained CoSn($10\bar{1}0$) single-crystalline films with high anisotropy of $\rho$ between the *c*- and *a*-axes. Other times, however, we obtained CoSn films with poorer crystallinity and a different surface morphology, as shown in Fig. S5. We later found that depositing a homoepitaxial MgO buffer layer solved the reproduction issue.

As shown in Fig. S5(a), this sample exhibited a strong CoSn $20\bar{2}3$ peak, in addition to the $10\bar{1}0$ peak and its higher-order reflections, indicating poorer CoSn($10\bar{1}0$) crystallinity than the samples with a homoepitaxial MgO buffer layer, as shown in Fig. 3(a) [$T_{\rm dep} \geq 350$ °C]. Figure S5(b) shows the surface morphology of this film. The grain-like surface morphology is longer along the *a*-axis in contrast to that shown in Fig. 3(f). The resistivity of the CoSn film was $\rho_{\parallel c}$ = 90.7 μΩ cm and $\rho_{\parallel a}$ = 173.0 μΩ cm, which still shows anisotropy between the *c*- and *a*-axes.

Although we cannot quantitatively separate the contributions of the intrinsic resistivity anisotropy and the surface morphology anisotropy of CoSn, this result indicates that the resistivity anisotropy observed in the CoSn($10\bar{1}0$) single-crystalline films is due not only to the surface morphology anisotropy but also due to the intrinsic resistivity anisotropy of CoSn.

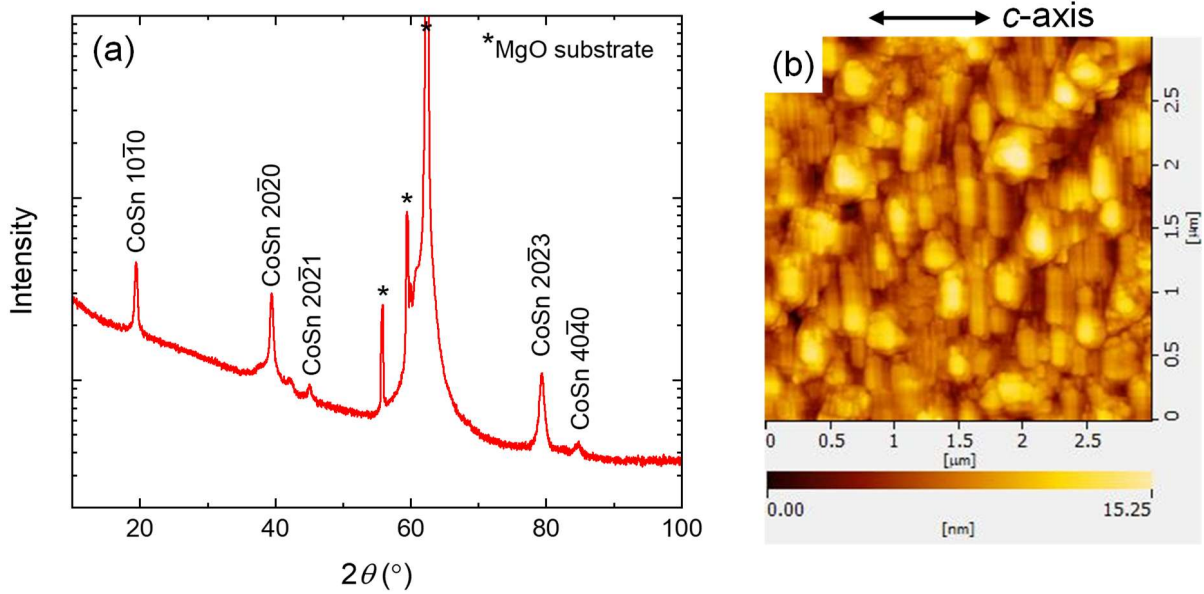

**FIG. S5.** (a) XRD profile and (b) AFM image of MgO(110) substrate/CoFe (2 nm)/Co (2 nm)/Ru (2 nm)/CoSn (30 nm) ($T_{dep}$ = 400 °C). The absence of an MgO (10 nm) homoepitaxial buffer layer resulted in a poorer CoSn($10\bar{1}0$) crystallinity and a different surface morphology compared to those with an MgO buffer layers (Fig. 3).

## 6. Separate EDS mapping images for each element and low magnification STEM image

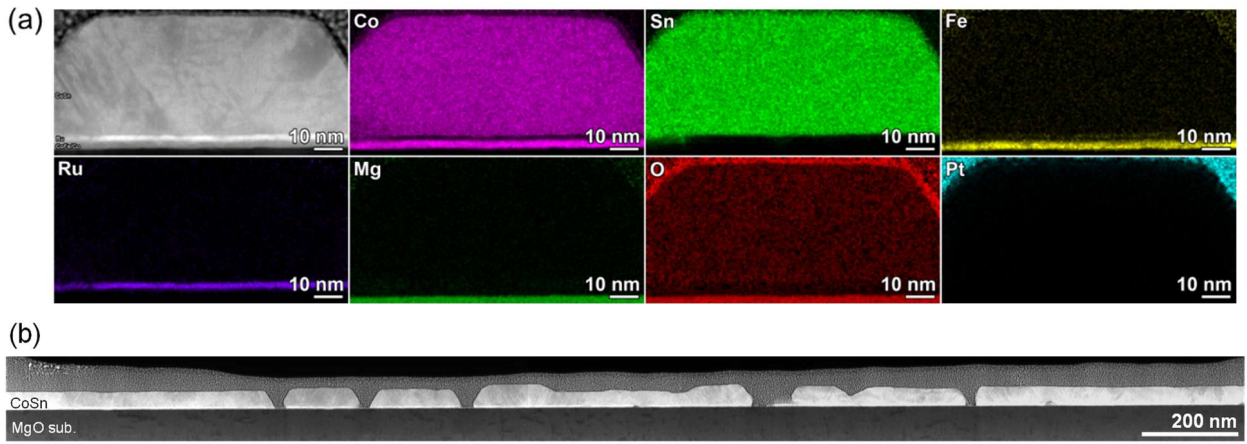

**FIG. S6.** (a) EDS elemental maps and (b) low magnification HAADF-STEM image of MgO(110) substrate/MgO (10 nm)/CoFe (2 nm)/Co (2 nm)/Ru (2 nm)/CoSn (30 nm) at $T_{dep}$ = 400 °C.